\documentclass[aps,prd,twocolumn]{revtex4-1}
\usepackage{amsmath,amssymb,amsthm,graphicx,latexsym, epsf}

\newcommand{\be}{\begin{equation}}
\newcommand{\ee}{\end{equation}}
\newcommand{\bea}{\begin{eqnarray}}
\newcommand{\eea}{\end{eqnarray}}
\newcommand{\bml}{\begin{subequations}}
\newcommand{\eml}{\end{subequations}}
\newcommand{\bfig}{\begin{figure}}
\newcommand{\efig}{\end{figure}}

\newcommand{\Del}{\Delta}

\begin{document}

\title{Fourth level MSSM inflation from new flat directions}

\author{Sayantan Choudhury$^{1}$\footnote{Electronic address: {sayanphysicsisi@gmail.com}} ${}^{}$
and Supratik Pal$^{1, 2}$\footnote{Electronic address: {supratik@isical.ac.in}} ${}^{}$}
\affiliation{$^1$Physics and Applied Mathematics Unit, Indian Statistical Institute, 203 B.T. Road, Kolkata 700 108, India \\
$^2$Bethe Center for Theoretical Physics and Physikalisches Institut der Universit\"{a}t Bonn, Nussallee 12, 53115 Bonn, Germany}



\begin{abstract}

We propose a model of inflation driven by minimal extension of SUSY, commonly known as MSSM. Starting from gauge invariant flat
directions in the $n=4$ level comprising of \textbf{QQQL,QuQd,QuLe} and \textbf{uude},
 we construct the inflaton potential and
employ it to investigate for its consequences  around the saddle point arising from the non-vanishing fourth derivative of the original potential. To this end, we derive the expressions for the important
parameters in MSSM inflation using the loop corrected
potential. We further  estimate the observable parameters
and find them to fit well with  recent observational data from WMAP7 by using the code CAMB. We
also explore the possibility of  primordial black hole  formation from our model.
Finally, we analyze one loop RGE and compute different phenomenological
parameters  which could be precisely determined in LHC or  future Linear Colliders.

\end{abstract}

\maketitle


\section{\bf Introduction}

The paradigm of primordial inflation is, by far, the most satisfactory explanation 
for early universe phenomena \cite{rocher}. As a general prescription, inflation occurs
due to a slowly rolling scalar field, the inflaton,
dynamically giving rise to an epoch of accelerated
expansion dominated by a false vacuum \cite{barrow}.
 Primordial quantum fluctuations of inflaton are responsible for creation of matter
content and observed perturbations in the Cosmic Microwave Background Radiation (CMBR).
Further, slow-roll inflationary scenario generically predicts almost Gaussian adiabatic perturbations with a nearly flat
spectrum, which conforms well with the latest observations.

Recently, some interesting proposition of inflationary model building was brought forth by Minimally Supersymmetric Standard Model (MSSM)
where the inflaton is a gauge invariant \cite{enqvist,arindam}
$n=4$ level combination of scalar superpartners squark and slepton
 fields and fermionic superpartner gauginos which are
 candidate Cold Dark Matter (CDM) particles. However the original potential for $n=4$ level
is unable to extract a suitable symmetry along the flat direction. To serve this purpose
the usual way is to incorporate {\it saddle point mechanism} to the
MSSM potential leading to  vanishing of the second derivative
 and the slow roll phase is driven by the next leading order derivative
 of the potential \cite{rocher,enqvist,arindam,maz,maz1}.
In most of the phenomenological situations, a fine tuning mechanism
is needed to place the flat direction field to the immediate neighborhood of the {\it saddle point}.
It is worthwhile to mention that MSSM inflation occurs at a comparative
 lower scale. This is in strong contrast with the
conventional class of models where the unfamiliar inflaton couplings to Standard Model (SM)
 are originated through arbitrary gauge singlets
 leading to the field magnitudes at GUT scale or higher, and hence,
 face problems in satisfactory quantitative estimation of
 a huge sector of the post-inflationary evolution i.e. thermal history
 of the early Universe, baryon
asymmetry and CDM. Herein lies the most appealing feature of MSSM inflation for
 which known SM couplings are measurable in
laboratory experiments such as Large Hadron Collider (LHC) \cite{lhc} or
 future linear colliders.

 In the present article we will consider a specific MSSM scenario where, for a specific choice of
soft supersymmetry (SUSY) breaking parameters A (trilinear couplings) and the inflaton mass $m_{\phi}$,
 the potential is D-flat along the
$\textbf{QQQL,QuQd,QuLe}$ and $\textbf{uude}$
directions. For our model existence of {\it saddle point} is guaranteed by the non-vanishing fourth derivative
of the potential, which makes the potential more flat than the previous ones.
This implies more precise information in the RG flow.
As we will show, this is the highest level of precision constraint one can impose
 on RG flow keeping the effective potential renormalizable
 in the vicinity of the {\it saddle point}. Our primary intention is to investigate for the analytical as well
as the numerical expressions for different observational parameters
  for MSSM inflation with these new flat directions. As it will turn out, they match quite well with latest observational
data from WMAP7 \cite{wmap07} and are expected to fit well with
upcoming data from PLANCK \cite{planck}. Additionally we have explicitly shown the connection between
running and running of the running of spectral index to the Primordial Black Hole (PBH) formation.
To this end we get the fine tuned parameter space
which is also in
good agreement with present estimates of cosmological frameworks.
We have further explored features of the MSSM from the solution of one
loop RGE which could be measured by LHC or future linear collider.

\section{\bf Flat directions and potential around saddle point }

Let us start with $n=4$ level superpotential \cite{martin}
\be\begin{array}{ll}\label{mssm}
 W^{nr}_{4}  = \frac{1}{M}\left[\sum_{I=1}^{24}\alpha_I  ({\bf QQQL})_I+
\sum_{I=1}^{81}\beta_I ({\bf QuQd})_I+\right.\\ \left.
~~~~~~~~~~~~\sum_{I=1}^{81}\gamma_I ({\bf QuLe})_I
+\sum_{I=1}^{27}\delta_I ({\bf uude})_I\right]
;\end{array}\ee

The renormalizable flat directions of the MSSM at n=4 level correspond to the gauge invariant
monomials subject to the four additional complex constraints \cite{martin} two each from
\be F^{\alpha}_{H_{u}}=\mu H^{\alpha}_{d}+\lambda^{ab}_{U}Q^{\alpha}_{a}u_{b}=0,\ee
\be F^{\alpha}_{H_{d}}=-\mu H^{\alpha}_{u}+\lambda^{ab}_{D}Q^{\alpha}_{a}d_{b}+\lambda^{ab}_{E}L^{\alpha}_{a}e_{b}=0,\ee
which can lift the flat directions which do not contain a Higgs field. Here $\lambda_{U}, \lambda_{D}$ and $\lambda_{E}$ are the Yukawa 
couplings, $H_{u}, H_{d}$ are the Higgs superfield and the $\mu$- term appears in the renormalizable part of the superpotential of MSSM . Consequently the equation(\ref{mssm}) 
breaks into four parts, each one of them now being flat:
\bea\label{xd1}W^{(1)}_{4}&=&
\frac{1}{M}\sum_{I=1}^{24}\alpha_I ({\bf QQQL})_I,
\\
\label{xd2} W^{(2)}_{4}&=&
\frac{1}{M}\sum_{I=1}^{81}\beta_I ({\bf QuQd})_I,
\\
\label{xd3} W^{(3)}_{4}&=&
\frac{1}{M}\sum_{I=1}^{81}\gamma_I ({\bf QuLe})_I,
\\
\label{xd4} W^{(4)}_{4}&=&
\frac{1}{M}\sum_{I=1}^{27}\delta_I ({\bf uude})_I,\eea

resulting in $W^{(i)}_{4}\approx \frac{\lambda_{4}}{4M}{\bf\Phi}^{4}  ~\forall i(=(1,2,3,4))$.
Considering any one of the above flat directions leads to the one loop corrected effective potential
\be\label{vg}
V(\phi,\theta)=\frac{1}{2}m^{2}_{\phi}\phi^{2}
+\frac{\lambda_{4}A}{4M}\phi^{4}Cos(4\theta+\theta_{A})
+\frac{\lambda^{2}_{4}}{M^{2}}\phi^{6},\ee

for all i. Here we define $\lambda_{4}=\lambda_{4,0}\left[1+D_{3}\log\left(\frac{\phi^{2}}{\mu^{2}_{0}}\right)\right]$,
 $A=\frac{A_{0}\left[1+D_{2}\log\left(\frac{\phi^{2}}{\mu^{2}_{0}
}\right)\right]}{\left[1+D_{3}\log\left(\frac{\phi^{2}}{\mu^{2}_{0}}\right)\right]}$ and
 $m^{2}_{\phi}=m^{2}_{0}\left[1+D_{1}\log\left(\frac{\phi^{2}}{\mu^{2}_{0}}\right)\right]$
 and in $\bf G_{MSSM}=SU(3)_{C}\otimes SU(2)_{L}\otimes U(1)_{Y}$ the representative flat direction field content is given by
\be\begin{array}{llll}\label{uip}
  {\bf Q^{I_{1}}_{a}}=\frac{1}{\sqrt{2}}({\bf\Phi},0)^{T},{\bf Q^{I_{2}}_{b}}=\frac{1}{\sqrt{2}}({\bf\Phi},0)^{T},{\bf Q^{I_{3}}_{c}}=\frac{1}{\sqrt{2}}({\bf\Phi},0)^{T},\\
~~~~~~~~ {\bf L^{I_{4}}_{3}}=\frac{1}{\sqrt{2}}P_{d}(0,{\bf\Phi})^{T},{\bf d^{B_{1}}_{a}}=\frac{{\bf\Phi}}{\sqrt{2}},
{\bf u^{B_{2}}_{b}}=\frac{{\bf\Phi}}{\sqrt{2}},\\~~~~~~~~~~~~~~~~~~~~~~ {\bf u^{B_{3}}_{c}}=\frac{{\bf\Phi}}{\sqrt{2}},{\bf e_{3}}=\frac{{\bf\Phi}}{\sqrt{2}}.
       \end{array}\ee

Here $m_{0},A_{0}$ and $\lambda_{4,0}$ are the values of the respective parameters at the scale $\mu_{0}$ and $D_{1},D_{2}$
and $D_{3}$ ($|D_{i}|\ll 1 \forall i$) are the fine tuning parameters. Additionally in the field contents 
${\bf 1 \leq B_{1},B_{2},B_{3} \leq 3}$ are color indices, ${\bf 1 \leq a,b,c\leq 3 }$ denote the indices for quark and lepton families and 
${\bf 1\leq I_{1},I_{2},I_{3},I_{4}\leq 2}$ are the weak isospin indices. The
flatness constraints require that ${\bf B_{1} \neq B_{2} \neq B_{3}}$ for quarks, ${\bf I_{1}\neq I_{2}\neq I_{3}\neq I_{4}}$, 
${\bf\sum^{3}_{d=1}P^{2}_{d}=1}$ ${\bf\forall P_{d}\in \mathbb{R}}$ for leptons and ${\bf a\neq b\neq c}$ for both. In eqn(\ref{vg}) $m_{\phi}$
represents the soft SUSY breaking mass term, $\phi$ the radial coordinate of the complex scalar field
 ${\bf\Phi}=\phi\exp(i\theta)$ ($\in\mathbb{C}$) and the second term is the so called
 A-term which has a periodicity of $2\pi$ in 2 D along with
an extra phase $\theta_{A}$. The radiative correction slightly affects the soft term and the value
of the saddle point.

 For $n=4$ we get an extremum for the principal values of $\theta$ at $\theta=\frac{(m\pi-\theta_{A})}{4}$
 (where ${\bf m\in \mathbb{Z}}$)

\be\begin{array}{ll}\label{sd} \phi_{0}=\sqrt{\frac{M}{4\lambda_{4}(3+D_{3})}}\left[A\left(1+\frac{D_{2}}{2}\right)\right.
\\ \left.\pm \sqrt{A^{2}\left(1+\frac{D_{2}}{2}\right)^{2}-8m^{2}_{\phi}(1+D_{1})(3+D_{3})}\right]^{\frac{1}{2}},
\end{array}
\ee

which appears from the constraint $V^{'}(\phi_{0})=0$ as a
necessary condition for {\it saddle point}. However, this condition alone will not lead to
{\it saddle point}. Rather, we have to make the potential sufficiently flat which can be achieved by vanishing 
higher derivatives of the potential. In this article, we consider non-vanishing fourth derivative of the potential 
resulting in saddle point. This will imply more fine-tuning but increased precision level in the information obtained 
from RG flow. Below we demonstrate how this is materialized.

 As discussed, $V^{''''}(\phi_{0})<0$ will give us secondary local minimum.
This leads to constraint relations:

\be\label{Ast}A =\sqrt{2(3+D_{3})G_{1}G_{2}G_{3}}m_{\phi}(\phi_{0}),\ee

\be\begin{array}{llll} \label{con2}  D_{3}=\frac{MA_{0}}{4\lambda_{4,0}\phi^{2}_{0}\left(37+60\log\left(\frac{\phi_{0}}{\mu_{0}}\right)\right)}
\left\{D_{2}\left(13+12\log\left(\frac{\phi_{0}}{\mu_{0}}\right)\right)\right.\\
\left.~~~~~~~~~~~~~~~~~~~~~~~~~~-\frac{2m^{2}_{\phi}(\phi_{0})D_{1}M}{\lambda_{4,0}A_{0}\phi^{2}_{0}}
+6\left(1-\frac{20\lambda_{4,0}\phi^{2}_{0}}{MA_{0}}\right)\right\}\end{array},\ee
one each for $V^{''}(\phi_{0})=0$ and $V^{'''}(\phi_{0})=0$. In this context $G_{1}=\left[\frac{(1+D_{1})}{(3+D_{3})}(15+11D_{3})-(1+3D_{1})\right]^{2}$,
$G_{2}=\left[(1+D_{1})\left(3+\frac{7}{2}D_{3}\right)-(1+3D_{1})\left(1+\frac{D_{2}}{2}\right)\right]^{-1}$,
$G_{3}=\left[\frac{\left(1+\frac{D_{2}}{2}\right)}{(3+D_{3})}(15+11D_{3})-\left(3+\frac{7}{2}D_{3}\right)\right]^{-1}$.
For the limit $|D_{1}|\ll 1$,$|D_{2}|\ll 1$ and $|D_{3}|\ll 1$ which gives
$\phi_{0}=\phi^{tree}_{0}\left[1+\frac{D_{1}}{2}-\frac{D_{3}}{6}\right]^{\frac{1}{2}}$ and
$A\simeq A_{tree}\left[1+\frac{D_{1}}{2}-\frac{D_{3}}{6}\right]$,
where $\phi^{tree}_{0}=\sqrt{\frac{m_{\phi}(\phi_{0})M}{\lambda_{4}\sqrt{6}}}$ and $A_{tree}=2\sqrt{6}m_{\phi}(\phi_{0})$
represents tree level expressions. This means, during RG flow mentioning two parameters only ($D_{1}$ and $D_{2}$) will
suffice instead of the usual three parameters in earlier MSSM models. This results in more precise information in RG flow. 
One may get tempted to vanish further higher derivatives of the potential in order to evaluate other
 unknown parameters ($D_{1}$ and $D_{2}$) without going into RG flow but this will make the effective inflaton potential
 in the vicinity of {\it saddle point} non-renormalizable. So, this is the highest level of precision 
constraint one can impose on RG flow parameters.

Consequently, around the {\it saddle point} $\phi_{0}$, the inflaton potential can be expanded in a Taylor series as,

\be\label{hgkl} V(\phi)=\tilde{C}_{0}+\tilde{C}_{4}(\phi-\phi_{0})^{4},\ee
where $\tilde{C}_{0}=V(\phi_{0})=\frac{m^{3}_{\phi}(\phi_{0})M}{6\sqrt{6}\lambda_{4}}
\left\{3\left(1+\frac{D_{1}}{2}-\frac{D_{3}}{6}\right)\right.\\ \left.\left[1+D_{1}\log\left(\frac{\phi^{2}_{0}}{\mu^{2}_{0}}\right)\right]-3\left(1+\frac{D_{1}}{2}-\frac{D_{3}}{6}\right)^{2}\left[1+D_{2}\log\left(\frac{\phi^{2}_{0}}{\mu^{2}_{0}}\right)\right]\right.
\\  \left.+\left(1+\frac{D_{1}}{2}-\frac{D_{3}}{6}\right)^{2}\left[1+D_{2}\log\left(\frac{\phi^{2}_{0}}{\mu^{2}_{0}}\right)\right]
\right\}$
and\\
$\tilde{C}_{4}=\frac{1}{{4!}}V^{''''}(\phi_{0})
=\frac{m^{2}_{\phi}(\phi_{0})}{24\sqrt{6}\phi^{2}_{0}}\left(1+\frac{D_{1}}{2}-\frac{D_{3}}{6}\right)
\\ \left\{\left\{\left[\left(\frac{360}{\sqrt{6}}-12\sqrt{6}\right)+(684D_{3}-50\sqrt{6}D_{2})\right]
\left(1+\frac{D_{1}}{2}-\frac{D_{3}}{6}\right)\right.\right.
\\  \left.\left.-\frac{2\sqrt{6}D_{1}}{\left(1+\frac{D_{1}}{2}-\frac{D_{3}}{6}\right)}\right\}
  +\left(1+\frac{D_{1}}{2}-\frac{D_{3}}{6}\right)\left(\frac{360D_{3}}{\sqrt{6}}
-12\sqrt{6}D_{2}\right)\right.\\ \left.\log\left(\frac{\phi^{2}_{0}}{\mu^{2}_{0}}\right)\right\}$.

In what follows we shall model MSSM inflation with the above potential.

\section{\bf Modeling MSSM inflation and parameter estimation}

 For brevity, let us introduce a change of parameter $\phi$ $\rightarrow$ $x=\phi-\phi_{0}$
 which represents the inflaton with shifted origin.
Using this new notation of field the slow roll parameters \cite{lyth} are given by,

\bea\label{df1}\tiny\epsilon_{v}(x)&=&
\frac{M^{2}}{2}\left(\frac{V^{'}}{V}\right)^{2}
=\frac{8\tilde{C}^{2}_{4}M^{2}x^{6}}
{(\tilde{C}_{0}+\tilde{C}_{4}x^{4})^{2}},
\\
\label{df2} \tiny\eta_{v}(x)&=&
 M^{2}\left(\frac{V^{''}}{V}\right)
=\frac{12\tilde{C}_{4}M^{2}x^{2}}
{(\tilde{C}_{0}+\tilde{C}_{4}x^{4})},
\\
\label{df3}\tiny\xi^{2}_{v}(x)&=&
M^{4}\left(\frac{V^{'}V^{'''}}{V^{2}}\right)
=\frac{96\tilde{C}^{2}_{4}M^{4}x^{4}}
{(\tilde{C}_{0}+\tilde{C}_{4}x^{4})^{2}},
\\
\label{df5}\tiny\sigma^{3}_{v}(x)&=&
M^{6}\left(\frac{(V^{'})^{2}V^{''''}}{V^{3}}\right)
=\frac{384\tilde{C}^{3}_{4}M^{6}x^{6}}{(\tilde{C}_{0}+\tilde{C}_{4}x^{4})^{3}},\eea
where a prime denotes $d/d\phi=d/dx$.
During slow-roll inflation $\epsilon_{v},|\eta_{v}|,|\xi_{v}|,|\sigma_{v}|\ll 1$ and the end of the inflation corresponds to
$|x_{f}|\simeq \left(\frac{\tilde{C}^{2}_{0}}{8M^{2}\tilde{C}^{2}_{4}}\right)^{\frac{1}{6}}$ where $x_{f}=\phi_{f}-\phi_{0}$.
In this context equation of state parameter can be expressed as
\be\begin{array}{lll}\label{state}
    \omega(x)=\frac{P(x)}{\rho(x)}
=\left[\frac{-\tilde{C}^{2}_{4}x^{8}+\frac{16}{3}\tilde{C}^{2}_{4}M^{2}x^{6}-2\tilde{C}_{0}
\tilde{C}_{4}x^{4}-\tilde{C}^{2}_{0}}{\tilde{C}^{2}_{4}x^{8}+\frac{16}{3}\tilde{C}^{2}_{4}M^{2}x^{6}+2\tilde{C}_{0}
\tilde{C}_{4}x^{4}+\tilde{C}^{2}_{0}}\right]
   \end{array}\ee
\begin{figure}[htb]
{\centerline{\includegraphics[width=7cm, height=5cm]{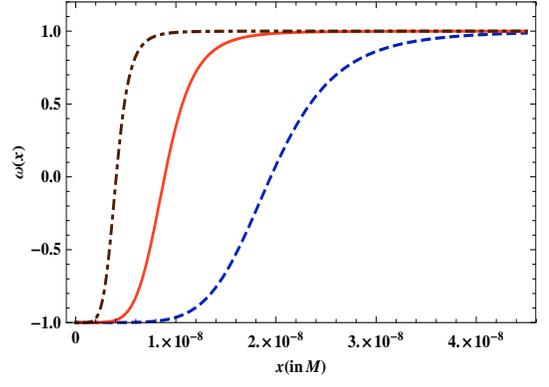}}}
\caption{Variation of equation of state parameter($\omega(x)$) versus shifted inflaton field ($x$)} \label{figVr38}
\end{figure}
 which implies the energy scale of MSSM inflation is $\tilde{C}^{\frac{1}{4}}_{0}\sim(0.409-1.301)\times 10^{-9}M$
explicitly shown in the allowed region in Fig(\ref{figVr38}).



 The number of e-foldings are defined \cite{lyth} for our model in the limit $\tilde{C}_{0}\gg\tilde{C}_{4}$ as
\be\begin{array}{ll}\label{uit}{\cal N}=\log\left(\frac{a(t_{f})}{a(t_{i})}\right)\simeq \frac{1}{M^{2}}\int^{x_{i}}_{x_{f}}\frac{V}{V^{'}}dx
\simeq\frac{\tilde{C}_{0}}{8M^{2}\tilde{C}_{4}}\left\{\frac{1}{x^{2}_{f}}-\frac{1}{x^{2}_{i}}\right\},\end{array}\ee

Further, in this framework the expressions for amplitude of the scalar perturbation, tensor perturbation
and tensor to scalar ratio are given by

\bea\label{x1}\Delta^{2}_{s}&=&
\frac{V^{3}}{75\pi^{2}M^{6}(V^{'})^{2}}|_{k=aH}
=-\frac{32\tilde{C}_{4}{\cal N}^{3}}{75\pi^{2}},
\\
\label{x2}\Delta^{2}_{t}&=&
\frac{V}{150\pi^{2}M^{4}}|_{k=aH}
=\frac{(\tilde{C}_{0}+\tilde{C}_{4}x^{4}_{\star})}{150\pi^{2}M^{4}},
\\
\label{x3} r &=&
16\left(\frac{\Delta^{2}_{t}}{\Delta^{2}_{s}}\right)
=-\frac{\tilde{C}_{0}}{4\tilde{C}_{4}M^{4}{\cal N}^{3}}.\eea

Here $x_{\star}$ represents the value of the inflaton field
at the horizon crossing. For our model expression for spectral index, running and running of the running reduces to the following form:
\bea
\label{d1}
n_{s}-1
&=&\frac{3\tilde{C}_{0}}{32\tilde{C}_{4}M^{4}{\cal N}^{3}}-\frac{3}{{\cal N}}
,\\
 \label{d11}n_{t}&=&
\frac{\tilde{C}_{0}}{32\tilde{C}_{4}M^{4}{\cal N}^{3}},\\ \label{famm}
\alpha_{s}
&=&\frac{3}{{\cal N}^{2}}-\frac{2}{3}(n_{s}-1)^{2},\\ \label{ramm}
\kappa_{s}
&=&-\frac{6}{{\cal N}^{3}}-\frac{4}{{\cal N}^{2}}(n_{s}-1)+\frac{8}{9}(n_{s}-1)^{3}.\eea

\begin{figure}[htb]
{\centerline{\includegraphics[width=7cm, height=5cm]{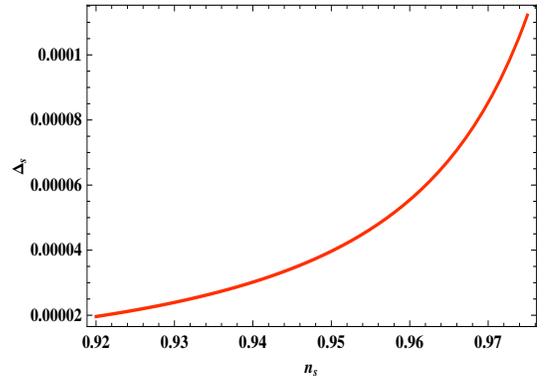}}}
\caption{$~$ Variation of the$~$
 scalar power spectrum($\Del_{s}$) vs scalar spectral index($n_{s}$)} \label{figVr8667}
\end{figure}

\begin{figure}[htb]
{\centerline{\includegraphics[width=7cm, height=5cm] {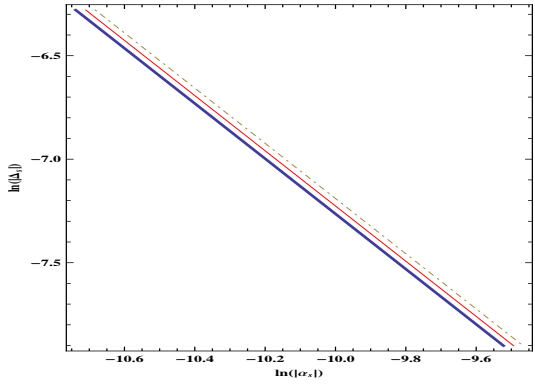}}}
\caption{$~$ Parametric plot of the logarithmic scaled amplitude of the scalar fluctuation ($ln({\Delta}_{s})$)
 vs logarithmic scaled amplitude
of the running of the spectral index ($\ln(|\alpha_{s}|)$)$~$ 
.} \label{figVr241}
\end{figure}

Fig(\ref{figVr8667}) depicts the behavior of the scalar power spectrum as a function of scalar spectral index.
For ${\cal N}=70$ the scalar spectral index is within the bounds of 
 {\it WMAP7+BAO+h0 data} for model {\it $\Lambda$CDM+sz+lens} \cite{wmap07}.

 Fig(\ref{figVr241}) shows the 
behavior of amplitude of scalar fluctuation as a function of 
running of the spectral index. For the best fit value of $\tilde{C}_{0}=2.867\times 10^{-36}M^{4}$,
$\tilde{C}_{4}=-1.685\times 10^{-13}$ and ${\cal N}=70$ the cosmological parameters obtained from our model is
$\Del^{2}_{s}=2.498\times 10^{-9}$, $\Del^{2}_{t}=1.936\times 10^{-39}$, $n_{s}=0.957$, $n_{t}=-1.550\times 10^{-30}$,
$r=1.240\times 10^{-29}$, $\alpha_{s}=-0.612\times 10^{-3}$, $\kappa_{s}=1.749\times 10^{-5}$.

\begin{figure}[htb]
{\centerline{\includegraphics[width=9cm, height=6cm]{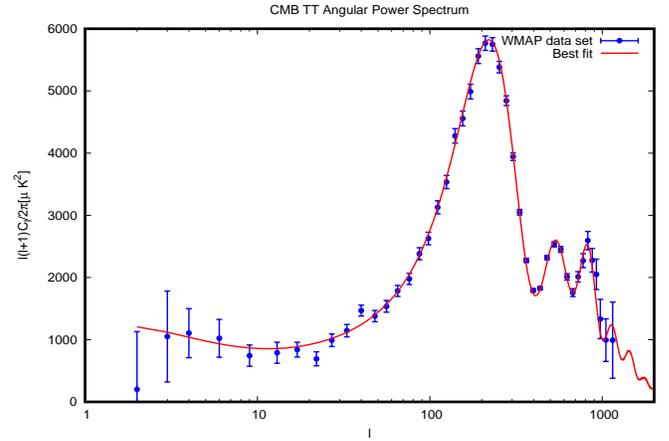}}}
\caption{Variation of CMB angular power spectrum$~$ $C^{TT}_{l}$ for
best fit and WMAP seven years data with the multipoles l for scalar mode} \label{figVr1181}
\end{figure}

Further, we use the publicly available code CAMB \cite{camb} to verify our results directly with observation.
To operate CAMB at the pivot scale $k_0=0.002~{\rm Mpc}^{-1}$ the values of the initial parameter space
are taken for $\tilde{C}_{0}=2.867\times10^{-36}M^{4}$ and ${\cal N}=70$.
 Additionally WMAP7 years dataset \cite{wmap07} for
$\Lambda$CDM background has been used in CAMB to obtain CMB angular power spectrum.
In Table\ref{tab3} we have given all the input parameters for CAMB.
Table\ref{tab4} shows the CAMB output, which is in good agreement with WMAP seven years data.
In Fig.\ref{figVr1181}
we have plotted CAMB output of CMB TT angular power spectrum
$C_l^{TT}$ for best fit with WMAP seven years data for scalar mode, which explicitly show
the agreement of our model with WMAP7 dataset.

\begin{table}[htb]
\begin{tabular}{|c|c|c|c|c|c|c|c|c|c|}
\hline $H_0$ & $\tau_{Reion}$ &$\Omega_b h^2$& $\Omega_c h^2
$& $T_{CMB}$
 \\
km/sec/MPc& & && K\\
 \hline
71.0&0.09&0.0226&0.1120&2.725\\
\hline
\end{tabular}
\caption{Input parameters}\label{tab3}
\end{table}
\begin{table}[htb]
\begin{tabular}{|c|c|c|c|c|c|c|c|c|c|}
\hline $t_0$ & $z_{Reion}$ &$\Omega_m$&$\Omega_{\Lambda}$&$\Omega_k$&$\eta_{Rec}$& $\eta_0$
 \\
Gyr& & && &Mpc & Mpc\\
 \hline
13.707&10.704&0.2670&0.7330&0.0&285.10&14345.1\\
\hline
\end{tabular}
\caption{Output obtained from CAMB}\label{tab4}
\end{table}

Now in the context of any running mass model one can expand the
spectral index with the following parameterization \cite{kosowsky}:
\be \label{ntwer}
 n({\cal R}) = n_z (k_0) - \frac{\alpha_z(k_0)}{2!}\ln\left( k_0
 {\cal R} \right) + \frac{\kappa_z(k_0)}{3!}\ln^2\left( k_0 {\cal R} \right)+.....
\ee
with ${\cal R} \ll 1/k_0$, i.e. $\ln(k_0 {\cal R}) < 0$. This is identified to be the significant
 contribution to the Primordial Black Hole (PBH) formation. Here the
parameterization index $ z:\left[s(scalar),t(tensor)\right]$ and the explicit form of
 the first term in the above expansion is given by
\be\label{runfunc}  n_z (k_0)=
\left\{
	\begin{array}{ll}
		 n_s (k_0)-1 & \mbox{if } z=s \\
		 n_t (k_0) & \mbox{if } z=t.
	\end{array}
\right.
\ee

Existence of the running and running of the running is
 the key feature in the formation of PBH in the radiation dominated era just after inflation \cite{dress}
which could form CDM in the Universe. The
initial PBHs mass $\cal M_{\text{PBH}}$ is related to the particle horizon mass
${\cal M}$ by
$ {\cal M_{\text{PBH}}}={\cal M }\gamma= \frac{4\pi}{3}\gamma \rho  {\cal H}^{-3}\,$
at horizon entry, ${\cal R}=(a{\cal H})^{-1}$. This is formed when the density
fluctuation exceeds the threshold for PBH formation given as in {\it Press--Schechter theory} by \cite{dress,Copeland}
\begin{equation} \label{fofm}
  f( \geq{\cal M}) = 2 \gamma \int _{\varTheta_{\rm th}}^{\infty} {\cal\text{d}\varTheta\,}{\cal P(\varTheta; M(R))}
.
\end{equation}
Here $\cal P(\varTheta; M(R))$ is the {\it Gaussian probability distribution function} of the
linearized density field $\varTheta$ smoothed on a comoving scale $\cal R$ by
${\cal P(\varTheta;R)} = \frac {1} {\sqrt{2\pi} \varSigma_{\varTheta}({\cal R})} \exp\left( -\frac
{\varTheta^{2}} {2\varSigma_{\varTheta}^{2}({\cal R})} \right)$
where the standard deviation
\begin{equation} \label{sigma}
\varSigma_{\varTheta}({\cal R}) =\sqrt{ \int_{0}^{\infty}\dfrac{\text{d}k} {k} \exp\left(-k^{2}{\cal R}^{2} \right)\ {\Delta}^2_{\varTheta}(k)
}\,.
\end{equation}

  For our model power spectrum for $\varTheta(k,\eta)$ is given by
\be\begin{array}{ll}\label{psx1}{\Delta}^2_{\varTheta}(k,\eta)
=\frac{4J} {25}\left( \frac{k}{a{\cal H}}
\right)^4 {\Delta}^2_{s}(k,\eta)=\frac{8J\sqrt{3\tilde{C}_{0}}Gk^4\eta^4(1+k^2\eta^2)} {25M\pi^2\kappa_{s}(\eta)},
\end{array}\ee
 where $\kappa_{s}(\eta)=\frac{18\sqrt{3}M^3G^3}{\tilde{C}^{\frac{3}{2}}_{0}\left[\Phi(f)-MG\sqrt{\frac{3}{
\tilde{C}_{0}}}\ln\left(\frac{\eta a(t_f)\sqrt{\tilde{C}_{0}}}{\sqrt{3}M}\right)\right]^3}$ is the equivalent
expression for running of the running in terms of conformal time $\eta$ and $J=\frac{(1+w)^2} {(1+\frac{3}{5}w)^2}$.
 Additionally we have used
$\Phi(f)=x^{-2}_{f}$ and $G=\frac{8\tilde{C}_{4}M}{\sqrt{3\tilde{C}_{0}}}$.
Substituting eqn(\ref{psx1}) in eqn(\ref{sigma}) and using eqn(\ref{ramm})
 at the horizon
crossing we get $\varSigma_{\varTheta}({\cal R})=\sqrt{\frac{8\sqrt{3\tilde{C}_{0}}JG}{25M\pi^2\kappa_{s}}\Gamma[\frac{(n_S({\cal R})+3)}{2}]}$ for
$n_s({\cal R})>3$. Consequently eqn(\ref{fofm}) gives
$f( \geq{\cal M}) =  \gamma\text{erf}\left[\frac{\varTheta_{\rm th}}
{\sqrt{2} \varSigma_{\varTheta}({\cal R})}\right]$. Here we fix $\gamma
\simeq 0.2$ during the radiation dominated era \cite{Carr2} for proper numerical estimations. In general
 the mass of PBHs is expected to depend
on the amplitude, size and shape of the perturbations \cite{Niemeyer}.
As a consequence the PBH mass is given by \cite{dress}
${\cal M_{\text{PBH}}} = \gamma {\cal M_{\text{eq}}} (k_{\text{eq}}{\cal R})^2 \left( \frac
{g_{\ast,\text{eq}}} {g_{\ast}} \right)^{\frac{1}{3}}\, ,$
where the subscript ``eq'' refers to the matter--radiation
equality. Here we use
$g_{\star}=228.75$ (all degrees of freedom
in MSSM), while $g_{\ast,\text{eq}} = 3.36$
and $k_{eq} = 0.07 \Omega_{\text{m}} h^{2}\ \text{Mpc}^{-1}$ (Here we use $\Omega_{\text{m}}
h^{2} = 0.2670$ from the CAMB output). Consequently
 the relation between comoving scale and the PBH mass in the context of MSSM is given by

\begin{equation} \label{R}
\frac {\cal R} {1\ \text{Mpc}} = 1.250 \times 10^{-23}\left(
\frac {\cal M_{\text{PBH}}} {1\ \text{g}} \right)^{\frac{1}{2}}\, .
\end{equation}

\begin{figure}[htb]
{\centerline{\includegraphics[width=7cm, height=5cm]{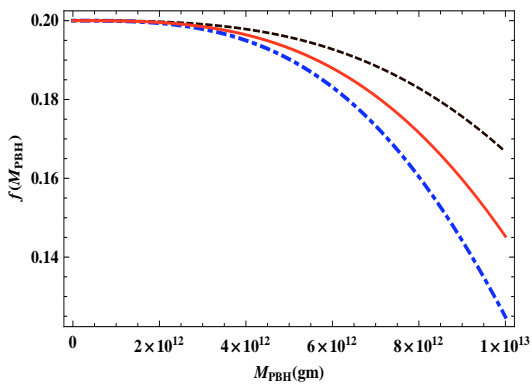}}}
\caption{$~$ Variation of the$~$
 fraction of the energy density of the universe collapsing into PBHs as a function
of the PBH mass, for three different values of the threshold
$\varTheta_{th} = 0.3(dashed), 0.5(solid), 0.7(dotdashed)$.
} \label{figVr8167}
\end{figure}

 Fig(\ref{figVr8167}) shows the behavior of {\it Press--Schechter mass function}
with respect to PBH mass. With the values of the parameters as obtained earlier, we have 
${\cal M}_{PBH}\simeq 10^{13}gm$ and
the corresponding fractional energy density $f=0.170$. 

Finally, the reheating temperature for our model turns out to be
$T_{rh}=\left(\frac{30x^{2}_{\star}M^{2}}{g_{\star}\pi^{2}}\right)^{\frac{1}{4}}
\sqrt[12]{1200\pi^{2}\tilde{C}^{2}_{4}\Del^{2}_{s}}$.
For ${\cal N}=70$ it is estimated as $T_{rh}=2.114\times 10^{8}GeV$ which is
 obviously significant input to choose the fine tuned initial conditions
for RGE flow discussed in the next section.

\section{One loop RG flow}

  For the flat direction \textbf{QQQL,QuQd,QuLe,uude} the soft SUSY breaking masses can be expressed as
\be\begin{array}{llll}\label{jkhjg}
  (m^{2}_{\phi})_{\textbf{QQQL}}=\frac{1}{4}(m^{2}_{{\bf\tilde{Q}_{a}}}
+m^{2}_{{\bf\tilde{Q}_{b}}}+m^{2}_{{\bf\tilde{Q}_{c}}}+m^{2}_{{\bf\tilde{L}_{3}}}),
\\ (m^{2}_{\phi})_{\textbf{QuQd}}=\frac{1}{4}(m^{2}_{{\bf\tilde{Q}_{a}}}
+m^{2}_{{\bf\tilde{Q}_{b}}}+m^{2}_{{\bf\tilde{u}_{c}}}+m^{2}_{{\bf\tilde{d}_{3}}}),
 \\  (m^{2}_{\phi})_{\textbf{QuLe}}=\frac{1}{4}(m^{2}_{{\bf\tilde{Q}_{a}}}
+m^{2}_{{\bf\tilde{u}_{b}}}+m^{2}_{{\bf\tilde{L}_{c}}}+m^{2}_{{\bf\tilde{e_{3}}}}),
\\  (m^{2}_{\phi})_{\textbf{uude}}=\frac{1}{4}(m^{2}_{{\bf\tilde{u}_{a}}}+m^{2}_{{\bf\tilde{u}_{b}}}
+m^{2}_{{\bf\tilde{d}_{c}}}+m^{2}_{{\bf\tilde{e}_{3}}}),
\end{array}
\ee
where ${\bf1\leq a,b,c\leq 3}$ and ${\bf a\neq b\neq c}$. After neglecting the contribution from the all Yukawa couplings except from the top we can express the one-loop beta function as
\cite{nilles}
$\beta_{m^{2}_{a}}=\dot{\mu}m^{2}_{a}
=\frac{1}{8\pi^2}\left(m^2_{a}+|A^{33}_{U}|^2\right)\left(\lambda^{33}_{U}\right)^{2}-\frac{1}{2\pi^{2}}\sum^{3}_{\alpha=1}g^{2}_{\alpha}|\tilde{m}_{\alpha}|^{2}\textbf{X}_{\alpha a}$
where $\textbf{X}_{\alpha a}$ are the quadratic Casimir Group Invariants for the superfield $\Phi$, defined in terms
of Lie Algebra generators $T^{a}$ by
$(T^{\alpha}T^{\alpha})^{a}_{b}=\textbf{X}_{\alpha a}\delta^{a}_{b}$
and $\dot{\mu}=\mu\frac{\partial}{\partial\mu}$.


\begin{figure}[htb]
{\centerline{\includegraphics[width=7cm, height=5cm]{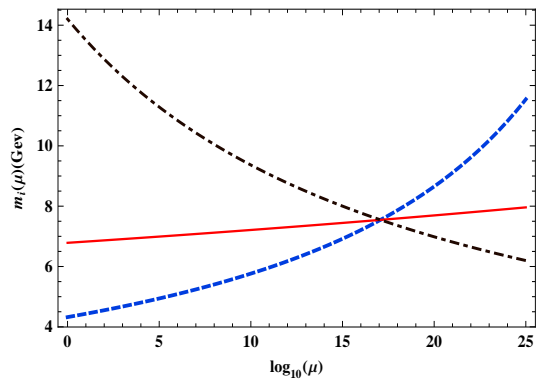}}}
\caption{Running of gaugino mass ($m_{i}(\mu)$)
 in one loop RGE for MSSM with the logarithmic scale $\log_{10}\left(\mu\right)$. Here we have used
$\mu_{0}=2.6\times 10^{7} GeV$ , $m_{i}(\mu_{0})=7.546\times10^{-3}TeV$, $\zeta=1$ $\forall$ i.} \label{figVr18}
\end{figure}

\begin{figure}[htb]
{\centerline{\includegraphics[width=8cm, height=5.5cm]{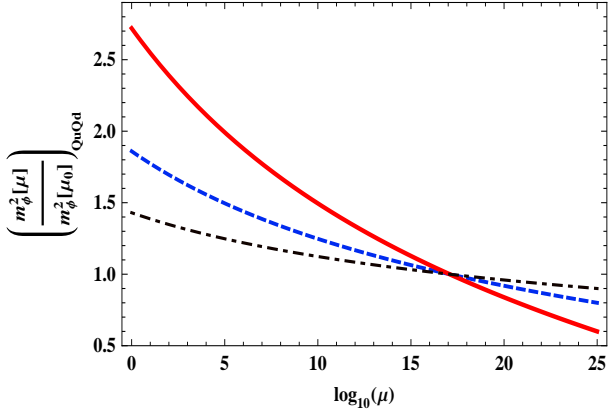}}}
\caption{Running of soft mass squared ratio $\left(\frac{m^{2}_{\phi}(\mu)}{m^{2}_{\phi}(\mu_{0})}\right)$ in one loop RGE for MSSM 
with the logarithmic scale $\log_{10}\left(\mu\right)$ where
$\mu_{0}=2.6\times 10^{7} GeV$ for $\zeta=0.5,1,2$ for $n=4$ level $~${\bf QuQd} $\forall$ i. Similar plots can be obtained for
{\bf QuLe}, {\bf QQQL} and
{\bf uude} flat directions also $\forall$ i.} \label{figVr1568}
\end{figure}
\begin{figure*}
{\centerline{\includegraphics[width=6.4cm, height=5cm]{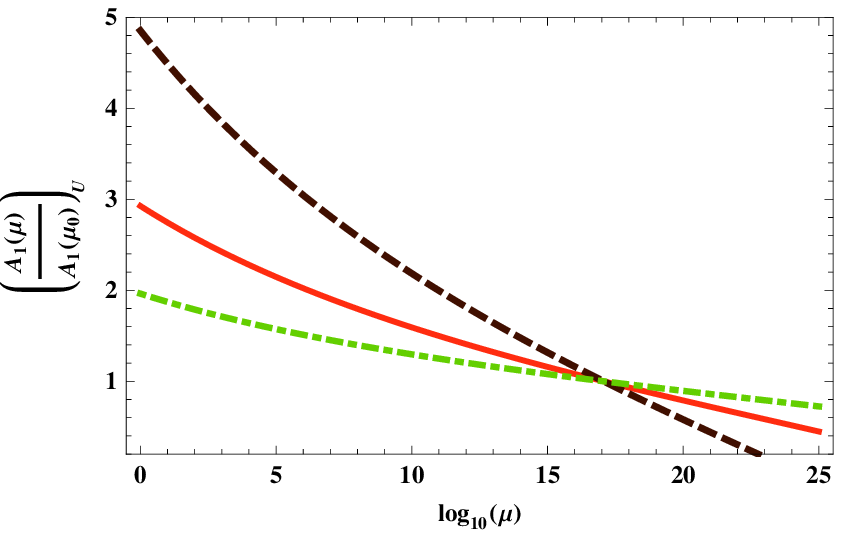} \includegraphics[width=6.4cm, height=5cm]{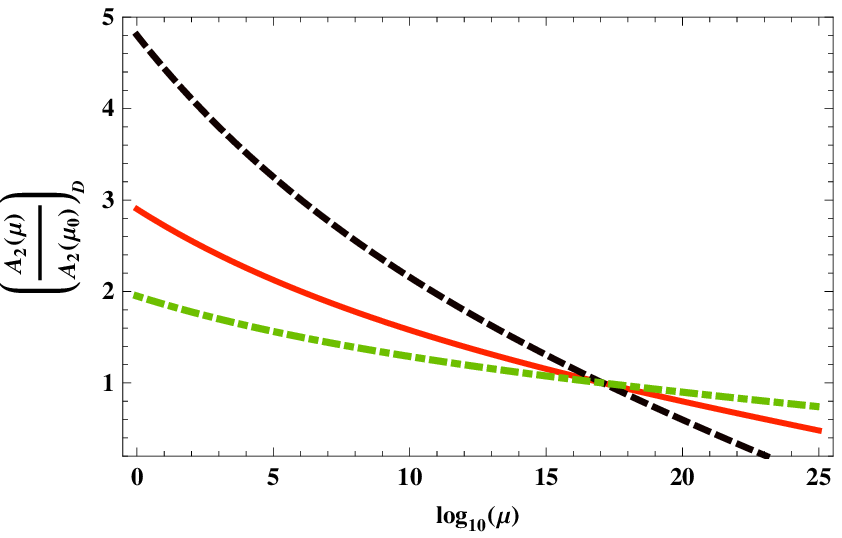} \includegraphics[width=6.4cm, height=5cm]{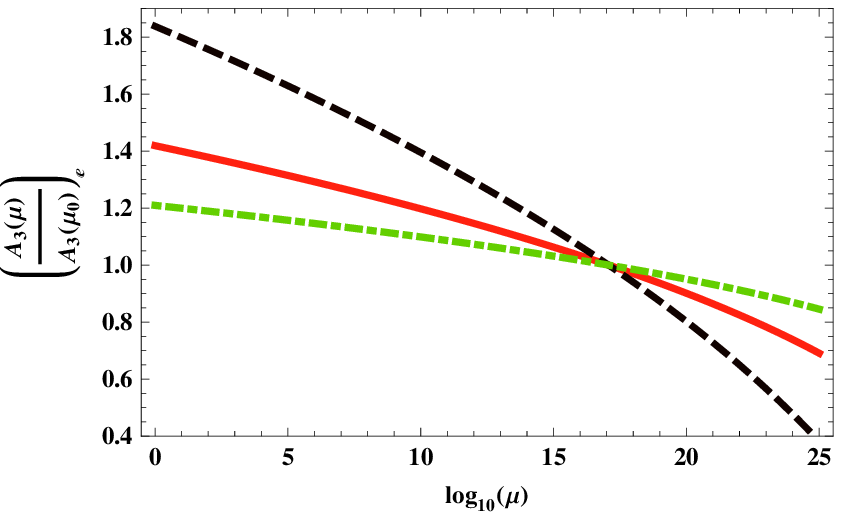}}}
\caption{Running of trilinear A -term ratio $\left(\frac{A_{\beta}(\mu)}{A_{\beta}(\mu_{0})}\right)$ in one loop RGE for MSSM with the logarithmic scale $\log_{10}\left(\mu\right)$ where
$\mu_{0}=2.6\times 10^{7} GeV$, $\zeta=0.5({\bf dotdashed}),
1({\bf solid}),2({\bf dashed})$ and $\beta=1(U),2(D),3(E)$ for $n=4$ level $\forall$ i.} \label{figVr118}
\end{figure*}

In the context of MSSM \\ $\textbf{X}_{1a}=\frac{3{\bf Y}^{2}_{a}}{5}$(for each ${\bf\Phi}_{a}$ with weak hyper charge ${\bf Y}_{a}$),
\\$\textbf{X}_{2a}=\frac{3}{4}$ (for ${\bf\Phi}_{a}={\bf Q,L,H_{u},H_{d}}$),
\\  $~~~~~~= 0 $ (for ${\bf\Phi}_{a}={\bf \bar{u},\bar{d},\bar{e}}$),
\\$\textbf{X}_{3a}=\frac{4}{3}$ (for ${\bf\Phi}_{a}={\bf Q,\bar{u},\bar{d}}$),
\\  $~~~~~~= 0 $ (for ${\bf\Phi}_{a}={\bf L,\bar{e},H_{u},H_{d}}$),
 \\where $\textbf{X}_{1a}$, $\textbf{X}_{2a}$ and $\textbf{X}_{3a}$ are applicable for ${\bf U(1)_{Y}}$,${\bf SU(2)_{L}}$ and
${\bf SU(3)_{C}}$ respectively.
 So for the flat direction content \textbf{QQQL,QuQd,QuLe,uude} we have the following beta functions:\\ \newpage
(a) \textbf{For Soft mass}:\\
    $\dot{\mu}(m^{2}_{\phi})_{\textbf{QQQL}}= \frac{1}{8\pi^2}\left(3m^{2}_{Q_{3}}+m^{2}_{U_3}+|A^{33}_{U}|^2\right)\left(\lambda^{33}_{U}\right)^{2}$
\\$~~~~~~~~~~~~~~~~~~~-\frac{1}{8\pi^{2}}\left(3g^{2}_{2}|\tilde{m}_{2}|^{2}+4g^{2}_{3}|\tilde{m}_{3}|^{2}\right),$
\\
$\dot{\mu}(m^{2}_{\phi})_{\textbf{QuQd}}=\frac{1}{2\pi^2}\left(m^{2}_{Q_{3}}+m^{2}_{U_3}\right)\left(\lambda^{33}_{U}\right)^{2}$\\
$~~~~~~~~~~~~~~~~~~~~~~-\frac{1}{8\pi^{2}}\left(\frac{3}{2}g^{2}_{2}|\tilde{m}_{2}|^{2}+\frac{16}{3}g^{2}_{3}|\tilde{m}_{3}|^{2}\right),$
\\
$\dot{\mu}(m^{2}_{\phi})_{\textbf{QuLe}}=\frac{3}{8\pi^2}\left(m^{2}_{Q_3}+m^{2}_{U_3}\right)\left(\lambda^{33}_{U}\right)^{2}$\\
$~~~~~~~~~~~~~~~~~~~~~~-\frac{1}{8\pi^{2}}\left(\frac{3}{2}g^{2}_{2}|\tilde{m}_{2}|^{2}+\frac{8}{3}g^{2}_{3}|\tilde{m}_{3}|^{2}\right),$
\\
$\dot{\mu}(m^{2}_{\phi})_{\textbf{uude}}=\frac{1}{2\pi^2}\left(m^{2}_{Q_{3}}+m^{2}_{U_3}\right)\left(\lambda^{33}_{U}\right)^{2}
-\frac{1}{2\pi^{2}}g^{2}_{3}|\tilde{m}_{3}|^{2},$\\
(b) \textbf{For Trilinear A- term}:\\
 $~~~~~~\dot{\mu}A^{aa}_{D}=\delta_{b3}\left(\lambda^{33}_{U}\right)^{2}\frac{A^{33}_{U}}{8\pi^{2}}\\
~~~~~~~~~~~~~~~-\frac{1}{4\pi^{2}}\left(\frac{7}{18}g^{2}_{1}|\tilde{m}_{1}|^{2}+
\frac{3}{2}g^{2}_{2}|\tilde{m}_{2}|^{2}+\frac{8}{3}g^{2}_{3}|\tilde{m}_{3}|^{2}\right),$
\\
$~~~~~~\dot{\mu}A^{ab}_{U}=\frac{3(1+\delta_{a3})A^{33}_{U}}{8\pi^{2}}\left(\lambda^{33}_{U}\right)^{2}\\
~~~~~~~~~~~~~~~-\frac{1}{4\pi^{2}}\left(\frac{13}{18}g^{2}_{1}|\tilde{m}_{1}|^{2}+\frac{3}{2}g^{2}_{2}|
\tilde{m}_{2}|^{2}+\frac{8}{3}g^{2}_{3}|\tilde{m}_{3}|^{2}\right),$
\\
$~~~~~~\dot{\mu}A^{aa}_{E}=
-\frac{1}{4\pi^{2}}\left(\frac{3}{2}g^{2}_{1}|\tilde{m}_{1}|^{2}+\frac{3}{2}g^{2}_{2}|\tilde{m}_{2}|^{2}\right),$\\
(c) \textbf{For Fourth level Yukawa coupling}:\\
$~~~~~~\dot{\mu}\lambda^{aa}_{U}=\frac{3(1+\delta_{a3})}{8\pi^{2}}\left(\lambda^{33}_{U}\right)^{3}
-\frac{\lambda^{aa}_{U}}{4\pi^{2}}\left(\frac{13}{18}g^{2}_{1}+\frac{3}{2}g^{2}_{2}+\frac{8}{3}g^{2}_{3}\right),$
\\
$~~~~~~\dot{\mu}\lambda^{ab}_{D}=\delta_{b3}\left(\lambda^{33}_{U}\right)^{2}\frac{\lambda^{ab}_{D}}{8\pi^{2}}
-\frac{\lambda^{ab}_{D}}{4\pi^{2}}\left(\frac{7}{18}g^{2}_{1}+\frac{3}{2}g^{2}_{2}+\frac{8}{3}g^{2}_{3}\right),$
\\
$~~~~~~\dot{\mu}\lambda^{aa}_{E}=
-\frac{\lambda^{aa}_{E}}{4\pi^{2}}\left(\frac{3}{2}g^{2}_{2}+\frac{3}{2}g^{2}_{3}\right)$\\
where all the superscript a and b represent generation or family indices run from 1 to 3 physically representing the
first, second and third generation respectively. For the one-loop renormalization of gauge couplings and gaugino masses, one has in general
\be\begin{array}{llll}\label{poi}\beta_{g_{\alpha}}=\dot{\mu}g_{\alpha}=\frac{g^{3}_{\alpha}}{16\pi^{2}}\left[\Sigma_{a}\textbf{I}_{\alpha a}-3\textbf{X}_{\alpha G}\right],\\
  \beta_{m_{\alpha}}=\dot{\mu}m_{\alpha}=\frac{g^{2}_{\alpha}m_{\alpha}}{8\pi^{2}}\left[\Sigma_{a}\textbf{I}_{\alpha a}-3\textbf{X}_{\alpha G}\right]  
   \end{array}
\ee
where $\textbf{X}_{\alpha G}$ quadratic Casimir invariant of the group $[$ 0 for $\bf{U(1)}$ and $\bf{N}$ for $\bf{SU(N)}$ $]$, $\textbf{I}_{\alpha a}$ is the Dynkin 
index of the chiral supermultiplet $\bf{\Phi}_{a}$ $[$ normalized to $\frac{1}{2}$ for each fundamental representation of $\bf{SU(N)}$ and to 
$3\bf{Y}^{2}_{a}/5$ for $\bf{U(1)}_{Y}$ $]$. For the above mentioned flat direction
the running of gauge couplings ($g_{i}(\mu)$) and gaugino masses ($m_{i}(\mu)$) obey \cite{nilles},
$\dot{\mu}g_{i}=
\frac{d_{i}}{2}g^{3}_{i},~~~~~~
\dot{\mu}\left(\frac{m_{i}}{g^{2}_{i}}\right)=
0~\forall~~ i$
where for $i=1({\bf U(1)_{Y}}),2({\bf SU(2)_{L}}),3({\bf SU(3)_{C}})$ here $d_{1}=\frac{11}{8\pi^{2}}$,$d_{2}=\frac{1}{8\pi^{2}}$,$d_{3}=-\frac{3}{8\pi^{2}}$ 
which is the simpler version of the equation(\ref{poi}).
Now to show explicitly that the contributions from the top Yukawa coupling ($\lambda^{33}_{U}$) are very small
for an induced electroweak group $\textbf{G}_{EW}$=$\textbf{SU(2)}_{L}$ $\otimes$ $\textbf{U(1)}_{Y}$ breakdown,
let us start with the Higgs potential \cite{nilles,martin}
\be
\begin{array}{lllll}\label{yuki}V_{ Higgs}({\bf H,\bar{H}})=m^2_{1}|{\bf H}|^2+m^2_{2}|{\bf\bar{H}}|^2+m^{2}_{3}
\left({\bf H{\bar{H}}}+{\bf H^{\dagger}\bar{H^{\dagger}}}\right)\\
~~~~~~~~~~~~~~~~~~~~~~~~~~~~~+\frac{1}{8}\left(g^2_{1}+g^2_{2}\right)\left[|{\bf H}|^2-|{\bf\bar{ H}}|^2\right]^2,\end{array}
\ee
where  ${\bf H}=H_{u}$ and ${\bf \bar{H}}=H_{d}$ represent the Higgs superfields and
 the relative vev of the two
Higgses are given by 
\be\label{desy}
\begin{array}{lllllll}
 v=\sqrt{{\langle {\bf H}\rangle}^2+{\langle {\bf {\bar H}}\rangle}^2}
\\~~\displaystyle=\sqrt{\frac{2\left[m^2_{1}-m^2_{2}-\left(m^2_{1}+m^2_{2}\right)cos(2{\bf\theta})\right]}{\left(g^2_{1}+g^2_{2}\right)cos(2{\bf\theta})}}
\end{array}
\ee
with $tan(\theta)=\frac{{\langle {\bf {\bar H}}\rangle}}{{\langle {\bf H}\rangle}}$. Here $\theta$ represents 
an angular parameter which parameterizes MSSM. For the sake of convenience 
let us now write $cos(2\theta)$ appearing in equation(\ref{desy}) introducing new parameterization as \cite{iban}
$cos(2\theta)=\frac{w^2-1}{w^2+1}$ where $w=\frac{\frac{\langle {\bf H}\rangle}{v}}{\sqrt{1-\left(\frac{\langle {\bf H}\rangle}{v}\right)^2}}$.  
Consequently the top Yukawa coupling can be expressed as $\lambda^{33}_{U}=\frac{m_{U}}{v sin(\theta)}$ where $0\leq\theta<\frac{\pi}{2}$
and the top mass 43 GeV$\leq m_{U}\leq 170$ GeV$\ll\mu_{GUT}$ 
comes from the RG flow \cite{nilles}. It is evident from the above parameterization \cite{iban,kaku,nano} that
as $w\rightarrow 1$, $\theta\rightarrow \frac{\pi}{4}$ which implies ${\langle {\bf H}\rangle}$ and ${\langle {\bf\bar{ H}}\rangle}$ is very large and have the same order of magnitude. 
As a result the relative vev $v$ is also large and the top Yukawa coupling is very very small
for which one can easily neglect it from the RG flow at the energy scale of MSSM inflation as mentioned earlier.
The consequence of the large vev of Higgs field can be taken care of by introducing strongly interacting gauge group  ${\bf G_{NEW}= G_{S}\otimes SU(3)_{C}}$
and its superconformal version ${\bf G_{SCONF}=SU(3)_{SC}\otimes SU(3)_{C}}$ \cite{kobayashi}.


 In table(\ref{tab8}) we have tabulated the numerical values of vev of ${\bf H}$ and ${\bf \bar{H}}$, the angular parameter $\theta$,
tan($\theta$), $w$, the top mass $m_{U}$ and 
the top Yukawa coupling $\lambda^{33}_{U}$ contributing to the parameter space of MSSM for the $n=4$ level flat directions 
 \textbf{QQQL,QuQd,QuLe} and \textbf{uude}. It should be noted that appearance of large vev of Higgses as mentioned in table(\ref{tab8}) can easily be interpreted
when Einstein Hilbert term appears in the total action of the theory at lowest order approximation \cite{kalo} which
 is our present consideration. Consequently the contributions from the hard cutoff is sub-leading due to the soft conformal symmetry
breaking. This leads to small top Yukawa coupling in the restricted parametric space
 of MSSM characterized by the phenomenological bound:  43 $GeV\leq m_{U} \leq$ 170 $GeV$,  $1.006\leq tan(\beta)\leq  1.025$
for the n=4 flat directions.
 
 Neglecting all the sub-leading contributions arising from the top Yukawa coupling in the restricted parameter space of the MSSM, the solutions of these RGE for n=4 level flat directions can be written as

\be\begin{array}{llll}\label{s1}
g_{i}(\mu)=
\frac{g_{i}(\mu_{0})}{\sqrt{1-d_{i}g^{2}_{i}(\mu_{0})\ln\left(\frac{\mu}{\mu_{0}}\right)}},
\\
m_{i}(\mu)=m_{i}(\mu_{0})\left(\frac{g_{i}(\mu)}{g_{i}(\mu_{0})}\right)^{2},
\\
\Delta m^{2}_{\phi}=
\sum^{3}_{i=1}f_{F}^{i}\Delta m^{2}_{i},
\\
\Delta A^{ab}_{\beta}=
\frac{1}{2}\sum^{3}_{i=1}(C_{\beta}^{i})^{ab}\Delta m_{i},
\\
\lambda^{ab}_{\beta}(\mu)=
\lambda^{ab}_{\beta}(\mu_{0})\prod^{3}_{i=1}\left(\frac{g_{i}(\mu_{0})}{g_{i}(\mu)}\right)^{(C_{\beta}^{i})^{ab}},\end{array}\ee
Here $g_{i}(\mu_{0})$, $m_{i}(\mu_{0})$, $A_{\beta}(\mu_{0})$,  $m_{\phi}(\mu_{0})$ and  $\lambda_{\beta}(\mu_{0})$ represent the value of the gauge couplings, gaugino masses,
 trilinear couplings, soft SUSY braking masses and Yukawa couplings at the characteristic scale $\mu_{0}$. In equation(\ref{s1}) we
have used the following shorthand notations:

 $~~~~~~~~~~~~~\Delta A_{\beta}=A_{\beta}(\mu)-A_{\beta}(\mu_{0})$,\\ $~~~~~~~~~~~~~~~~~\Delta m_{i}=m_{i}(\mu_{0})-m_{i}(\mu)$,\\
$~~~~~~~~~~~~~~~~~\Delta m^{2}_{\phi}=m^{2}_{\phi}(\mu)-m^{2}_{\phi}(\mu_{0})$,\\ $~~~~~~~~~~~~~~~~~\Delta m^{2}_{i}=m^{2}_{i}(\mu_{0})-m^{2}_{i}(\mu)$,\\
 where the $\beta$ indices 1,2,3 represent U, D, E
respectively.

\begin{table}[htb]
\begin{tabular}{|c|c|c|c|c|}
\hline ${\bf f_{F}^{i}}$ & ${\bf i=1(U(1)_{Y})}$ &${\bf i=2(SU(2)_{L})}$&${\bf i=3(SU(3)_{C})}$\\
 \hline
F=1({\bf QQQL})&0&$\frac{3}{2}$&-$\frac{2}{3}$\\
\hline
F=2({\bf QuQd})&0&$\frac{3}{4}$&-$\frac{8}{9}$\\
\hline
F=3({\bf QuLe})&0&$\frac{3}{4}$&-$\frac{4}{9}$\\
\hline
F=4({\bf uude})&0&0&-$\frac{2}{3}$\\
\hline
\end{tabular}
\caption{Entries of $f_{F}^{i}$ matrix obtained from the solution of RGE}\label{tab5}
\end{table}

\begin{table}[htb]
\begin{tabular}{|c|c|c|c|c|}
\hline ${\bf (C_{\beta}^{i})^{ab}}$ & ${\bf i=1(U(1)_{Y})}$ &${\bf i=2(SU(2)_{L})}$&${\bf i=3(SU(3)_{C})}$\\
 \hline
$\beta$=1({\bf U}),a=b&$\frac{26}{99}$&6&-$\frac{32}{9}$\\
\hline
$\beta$=2({\bf D}),a=b&$\frac{14}{99}$&6&-$\frac{32}{9}$\\
\hline
$\beta$=3({\bf E}),$a\neq b$&$\frac{6}{11}$&6&0\\
\hline
\end{tabular}
\caption{Entries of $(C_{\beta}^{i})^{ab}$ matrix obtained from the solution of RGE}\label{tab6}
\end{table}
 
In equation (\ref{s1}) $f_{F}^{i}$ and $(C_{\beta}^{i})^{ab}$ are $(4\times 3)$ and $(3\times 3)$ matrices whose entries 
are tabulated in Table(\ref{tab5}) and Table(\ref{tab6}) respectively. It is obvious from the RGE that $\beta=1,2$ implies $a=b$ and $\beta=3$ implies $a\neq b$.

\begin{figure}[htb]
{\centerline{\includegraphics[width=7.5cm, height=5cm]{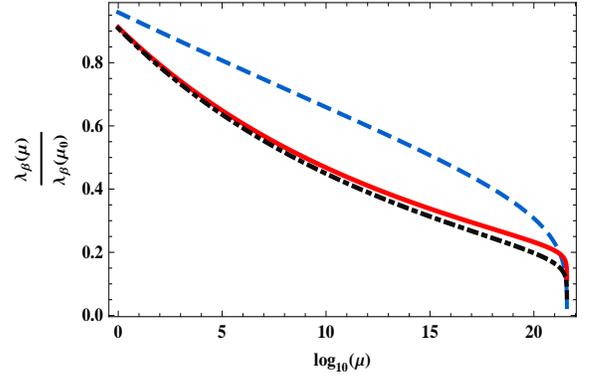}}}
\caption{Running of the ratio of the Yukawa coupling $\left(\frac{\lambda_{\beta}(\mu)}{\lambda_{\beta}(\mu_{0})}\right)$
 in one loop RGE for MSSM with the logarithmic scale $\log_{10}\left(\mu\right)$. Here we have used
$\mu_{0}=2.6\times 10^{7} GeV$ and $\beta=1(U),2(D),3(E)$ $\forall$ i.} \label{figVr178}
\end{figure}

Using the solutions of RGE along with the approximation that the running of the gaugino masses and gauge couplings
is very very small we get:

\be\begin{array}{lll}\label{cg1} D_{1}=
-\frac{1}{8\pi^{2}}\sum^{3}_{i=1}J_{i}\left(\frac{m_{i}}{m_{\phi_{0}}}\right)^{2}g^{2}_{i}(\mu_{0}),
\\
 D^{\beta}_{2}=
-\frac{1}{4\pi^{2}}\sum^{3}_{i=1}K^{\beta i}\left(\frac{m_{i}}{A_{0}}\right)g^{2}_{i}(\mu_{0}),\end{array}\ee

where we have $J_{1}=0$,$J_{2}=3$ and $J_{3}=4$ for $i=1, 2, 3$ and all the entries of  $K^{\beta i}$ $(3\times 3)$ matrix are tabulated in
table(\ref{tab7}).

\begin{table}[htb]
\begin{tabular}{|c|c|c|c|c|}
\hline ${\bf K^{\beta i}}$ & ${\bf i=1(U(1)_{Y})}$ &${\bf i=2(SU(2)_{L})}$&${\bf i=3(SU(3)_{C})}$\\
 \hline
$\beta$=1({\bf U})&$\frac{13}{18}$&$\frac{3}{2}$&$\frac{8}{3}$\\
\hline
$\beta$=2({\bf D})&$\frac{7}{18}$&$\frac{3}{2}$&$\frac{8}{3}$\\
\hline
$\beta$=3({\bf E})&$\frac{3}{2}$&$\frac{3}{2}$&0\\
\hline
\end{tabular}
\caption{Entries of $K^{\beta i}$ matrix}\label{tab7}
\end{table}
In this context the subscript `0' represents the values of parameters at the high scale $\mu_{0}$. As discussed in section III, constraining only $D_{1}$ and $D^{\beta}_{2}$ is sufficient here. Eqn(\ref{con2}) provides an 
extra constraint relation which restricts the parameters further leading to more precise information in RG flow.
For universal boundary conditions, the high scale is identified to be
the GUT scale $\mu_{GUT} \approx 3 \times
10^{16}$~GeV, ${\tilde m_{1}}(\mu_{GUT}) = {\tilde m_{2}}(\mu_{GUT}) ={\tilde m_{3}}(\mu_{GUT})= {\tilde m}$,
$A_{E}(\mu_{GUT})=A_{U}(\mu_{GUT})=A_{D}(\mu_{GUT})=A_{0}$
and $g_{1}\approx 0.56$, $g_{2}\approx
0.72$, $g_{3}\approx 0.85$. Now depending upon the different phenomenological situations the $n=4$ level flat directions 
are divided into two classes. The first class deals with ${\bf QuQd, QuLe}$
which is lifted completely at $n=4$ level. The other class which is lifted by higher dimensional operators
deals with ${\bf uude, QQQL}$. Most importantly ${\bf uude, QQQL}$ take part in the proton
 decay (${\it p\rightarrow\pi^{0}e^{+}}$, ${\it p\rightarrow\pi^{+}\nu_{e}}$ etc.) \cite{pdec} which introduces a stringent constraint
 on the Yukawa coupling $\lambda_{0}$ at $n=4$ level. Additionally the neutrino-antineutrino oscillation data
restricts $\lambda_{0}$ again.
Then we just use RG equations along with these restrictions to run the coupling constants and masses to the scales as mentioned in 
table(\ref{tab8}) with $M=2.4\times10^{18}$~GeV.

\begin{center}
\begin{table*}
{\small
\hfill{}
\begin{tabular}{|c|c|c|c|c|c|c|c|c|c|c|c|c|}
\hline ${\bf Flat~}$ & ${\bf\mu_{0}=\phi_{0}}$&$A_{0,tree} $&${\bf m_{\phi_{0}}}$&${\bf\langle {\bf H}\rangle }$& ${\bf\langle {\bf \bar{H}}\rangle }$&$\theta$&$tan(\theta)$&$w$&$v$&$m_{U}$ & ${\bf \lambda^{33}_{U}=\lambda_{0}}$\\
 ${\bf direction}$ & ${ GeV}$&${ GeV}$ &${ GeV}$&${ GeV}$&${ GeV}$&in $\deg$& &  &${ GeV}$&${ GeV}$&${ GeV}$\\
 \hline
${\bf QuLe}$&$2.6 \times
10^{7} $&$36.967$&$7.546$&$0.200\times 10^{16}$  & $0.458\times 10^{16}$& 45.171&1.006  &0.994  &$0.500\times 10^{16} $& 43&$1.212\times 10^{-14}  $\\
\hline
${\bf QuQd}$&$2.6 \times
10^{7} $&$36.967$&$7.546$&$ 0.450\times 10^{16} $ & $0.423\times 10^{16}$ & 45.370 &1.013  & 0.987 &$0.601\times 10^{16}$ &170 &$7.106\times 10^{-14}  $\\
\hline
${\bf QQQL}$&$1.344\times
10^{14}$&$892\times 10^{3}$&$182\times 10^{3}$&$0.188\times 10^{8}$&$0.124\times 10^{8}$&45.707 & 1.025 &0.975  &$0.226\times 10^{8}$& 80 &$4.945\times 10^{-6}$\\
\hline
${\bf uude}$&$2.896\times
10^{13}$&$4.142\times10^{6}$&$845\times 10^{3}$&$0.174\times 10^{6}$ &$0.157\times 10^{6}$ & 45.549 & 1.019 &0.981 &$0.235\times 10^{6}$ &135 & $8.047\times 10^{-4}$\\
\hline
\end{tabular}}
\hfill{}
\caption{MSSM parameter values obtained from RG flow for n=4 level flat directions}\label{tab8}
\end{table*}
\end{center}
Considering all these values we obtain effectively\\
$~~~~~~~~~~~~~~~~~~~~~D_{1} \approx  -0.056 \zeta^2, $\\
$~~~~~~~~~~~~~~~~~~~~~D^{1}_{2} \approx  -0.074 \zeta,$\\
$~~~~~~~~~~~~~~~~~~~~~D^{2}_{2} \approx  -0.071 \zeta,$\\
$~~~~~~~~~~~~~~~~~~~~~D^{3}_{2} \approx  -0.031 \zeta,$\\
$~~~~~~~~~~~~~~~~~~~~~D^{1}_{3}= D^{2}_{3}=D^{3}_{3}\approx  -0.048-0.168\zeta^{2}$,\\
where $\zeta = m/ m_{\phi}$ is calculated at the GUT scale. Typically the running
 based on gaugino loops alone results in negative values of
 $D_{i}\forall i$. Positive values can be obtained when one includes the Yukawa couplings,
practically the top Yukawa, but the order of magnitude remains the same. 
The choice of fine tuned initial conditions directly shows
 more fine tuning is required compared to other models. It is a straightforward exercise to verify that
even if one considers all the flat directions at $n=4$ level one will arrive at the potential eqn.(\ref{hgkl})
with same $\tilde{C}_{0}$ and $\tilde{C}_{4}$. This is precisely what we have done in this paper.

The results of RG flow have been demonstrated in figs(\ref{figVr18})-(\ref{figVr178}).
In fig(\ref{figVr18}) and fig(\ref{figVr1568})
`{\it dashed}', `{\it solid}' and `{\it dotdashed}' line represents ${\bf U(1)_{Y}}$, ${\bf SU(2)_{L}}$ and ${\bf SU(3)_{C}}$
gauge group content respectively. Fig(\ref{figVr18})-fig(\ref{figVr178}) explicitly showing
 the behavior of the RGE flow of gaugino masses, soft SUSY
breaking mass, trilinear couplings and Yukawa couplings respectively.
 Additionally fig(\ref{figVr18})-fig(\ref{figVr118}) give consistent GUT scale unification.

\section{Summary and outlook}

In this article we have proposed a model of inflation 
in the framework of MSSM with new flat directions using saddle point mechanism.
 We have demonstrated how we can construct the
 effective inflationary potential in the vicinity of the {\it saddle point} starting
 from $n=4$ level superpotential for the flat direction content
 \textbf{QQQL,QuQd,QuLe} and \textbf{uude} for MSSM. The effective inflaton potential
 around saddle point, resulting from the non-vanishing
fourth derivative of the original potential, has then been utilized in estimating for
 the observable parameters and confronting them with
WMAP7 dataset using the publicly available code CAMB, which reveals consistency of our model with latest observations. 
 We have then explored the possibility of Primordial Black Hole formation from the 
running-mass model by estimating the mass of PBH.

Subsequently, we have engaged ourselves in finding out the effective parameter space
and the constants appearing in the {\it saddle point} analysis for the MSSM inflation
 by solving the one loop RGE. It is worth mentioning that the RGE flow of fourth
level MSSM is exactly solvable in this context and we hope that
all the numerics can be tested in the LHC or any linear
collider in near future.
 Consequently we conclude
that fourth level MSSM inflation confronts extremely well
with WMAP7 within a certain parameter space obtained from one loop MSSM RGE flow.

A detailed survey of 
 RG flow with two loop beta function,
 inflection point inflation \cite{anupam3} for $n=4$ level
 MSSM candidates, sensitivity
 in the neighborhood of the saddle point with the one loop corrected
 potential, the effect of quantum Coleman De Luccia tunneling \cite{Coleman} and
 the inflationary model building of MSSM derived from string theory via braneworld using several compactification schemes remain an
open issue, which may even provide interesting signatures of MSSM
inflation. We hope to address some of these issues in due course.


\section*{Acknowledgments}

SC thanks H. P. Nilles, B. K. Pal and  I. Singh
for discussions and Council of Scientific and
Industrial Research, India for financial support through Junior
Research Fellowship (Grant No. 09/093(0132)/2010). SP is partially supported
by the Alexander von Humboldt Foundation Germany, the SFB-Tansregio TR33 ``The Dark Universe''
(Deutsche Forschungsgemeinschaft) and the European Union 7th
network program ``Unification in the LHC era''
(PITN-GA-2009-237920).
We also acknowledge illuminating discussions with A. Mazumdar which helped in improvement of the article.  




\end{document}